\pdfoutput=1
\documentclass[sigconf=true,review=false,natbib=true,anonymous=false,bookmarks=false]{acmart}
\usepackage{pifont}
\usepackage{multirow}%
\usepackage{tabularx}
\usepackage{enumitem}
\usepackage{tikz}
\usepackage{fontawesome5}
\usepackage{mwe}
\usepackage{subfig}
\usepackage{graphicx}
\usepackage{tabularx}   
\usepackage{hyperref}

\definecolor{loguipurple}{rgb}{0.274, 0.314, 0.612}
\newcommand{\cmark}{\ding{51}}%
\newcommand{\xmark}{\ding{55}}%

\newcommand{\textans}{\textcolor{textcolour}{\texttt{text}}}
\newcommand{\logui}{\color{loguipurple}\textbf{\texttt{Log}}\texttt{UI}\color{black}}
\newcommand{\voiceans}{\textcolor{voicecolour}{\texttt{voice}}}
\newcommand*\circled[1]{\tikz[baseline=(char.base)]{\node[shape=circle,fill=black,inner sep=1pt] (char) {\textcolor{white}{#1}};}}
\newcommand{\rqOne}{\textbf{RQ1}}
\newcommand{\rqTwo}{\textbf{RQ2}}
\newcommand{\rqThree}{\textbf{RQ3}}
\newcommand{\XSh}{\textbf{XS}}
\newcommand{\Sh}{\textbf{S}}
\newcommand{\M}{\textbf{M}}
\newcommand{\Lo}{\textbf{L}}
\newcommand{\XLo}{\textbf{XL}}
\newcommand{\SC}{\textbf{SC}}

\newcommand{\WM}{\textbf{WM}}
\newcommand{\In}{\textbf{IN}}

\definecolor{textcolour}{rgb}{0.2, 0.4, 1.0}
\definecolor{voicecolour}{rgb}{1.0, 0.4, 0.0}

\newcolumntype{Y}{>{\centering\arraybackslash}X}

\AtBeginDocument{%
  \providecommand\BibTeX{{%
    \normalfont B\kern-0.5em{\scshape i\kern-0.25em b}\kern-0.8em\TeX}}}
\usepackage{colortbl}

	\definecolor{lavendermist}{rgb}{0.9, 0.9, 0.98}
	\definecolor{aliceblue}{rgb}{0.94, 0.97, 1.0}
		\definecolor{antiquewhite}{rgb}{0.94, 1, 1}

\newcolumntype{D}[1]{>{\centering\arraybackslash}p{#1}}
\newcolumntype{L}[1]{>{\arraybackslash}p{#1}}
\copyrightyear{2023} 
\acmYear{2023} 
\setcopyright{rightsretained} 
\acmConference[SIGIR '23]{Proceedings of the 46th International ACM SIGIR Conference on Research and Development in Information Retrieval}{July 23--27, 2023}{Taipei, Taiwan}
\acmBooktitle{Proceedings of the 46th International ACM SIGIR Conference on Research and Development in Information Retrieval (SIGIR '23), July 23--27, 2023, Taipei, Taiwan}\acmDOI{10.1145/3539618.3591694}
\acmISBN{978-1-4503-9408-6/23/07}

\begin{document}
\fancyhead{}
    
\thanks{This research has been supported by \textit{NWO VIDI} project \textit{SearchX} (639.022.722) and \textit{NWO} project \textit{Aspasia} (015.013.027).}

    \title{Hear Me Out: A Study on the Use of the Voice Modality for Crowdsourced Relevance Assessments}

\author{Nirmal Roy}
\affiliation{
    \institution{Delft University of Technology}
    \country{The Netherlands}
}
\email{n.roy@tudelft.nl}

\author{Agathe Balayn}
\affiliation{
    \institution{Delft University of Technology}    
    \country{The Netherlands}
}
\email{a.m.a.balayn@tudelft.nl}

\author{David Maxwell}
\affiliation{
    \institution{Delft University of Technology}    
    \country{The Netherlands}
}
\email{maxwelld90@acm.org}
\author{Claudia Hauff}
\affiliation{
    \institution{Spotify \& Delft University of Technology}    
    \country{The Netherlands}
}
\email{c.hauff@tudelft.nl}


\begin{abstract}
The creation of relevance assessments by human assessors (often nowadays crowdworkers) is a vital step when building IR test collections. Prior works have investigated assessor quality \& behaviour, and tooling to support assessors in their task. We have few insights though into the impact of a document's \emph{presentation modality} on assessor efficiency and effectiveness. Given the rise of voice-based interfaces, we investigate whether it is feasible for assessors to judge the relevance of text documents via a voice-based interface. We ran a user study ($n=49$) on a crowdsourcing platform where participants judged the relevance of short and long documents---sampled from the \textit{TREC Deep Learning} corpus---presented to them either in the \textans{} or \voiceans{} modality. We found that: \textit{(i)} participants are \textit{equally} accurate in their judgements across both the \textans{} and \voiceans{} modality; 
\textit{(ii)} with increased document length it takes participants significantly longer (for documents of length $>120$ words it takes almost twice as much time) to make relevance judgements in the \voiceans{} condition; and
\textit{(iii)} the ability of assessors to ignore stimuli
that are not relevant (i.e., \textit{inhibition}) impacts the assessment quality in the \voiceans{} modality---assessors with higher inhibition are significantly more accurate than those with lower inhibition.
Our results indicate that we can reliably leverage the \voiceans{} modality as a means to effectively collect relevance labels from crowdworkers.
\end{abstract}



\maketitle

 \section{Introduction}\label{sec:intro}

Document relevance assessments by human assessors---with respect to a given set of \textit{information needs}---is a vital step in the building of an \textit{Information Retrieval (IR)} test collection~\cite{jones1976information,singhal2001modern}. 
Depending on the corpus, documents are represented in a variety of forms---including text (the most common form at \textit{TREC}), images~\cite{datta2008image,muller2010overview}, or videos~\cite{luan2011visiongo,gligorov2013evaluation}. Prior works have investigated assessor quality, their behaviour, and tooling to support assessors---most often in the context of text documents~\cite{scholer2011quantifying,scholer2013effect,anderton2013analysis,koopman2014relevation,roitero2021effect}. Given the prevalent nature of text corpora, we continue in this vein and focus on an aspect that has received little attention so far: the \emph{presentation modality} of the text documents during the judging process.


Thanks to the development of voice-based conversational search systems
, people have become accustomed to being presented search results that are read out to them, an approach that is very different from the presentation of text on-screen. 
We posit that by utilising such audio-based devices, we can increase the scope for collecting relevance judgements for text documents in a number of ways. For example, assessors can contribute by judging documents on their smartphones~\cite{vaish2014twitch,almeida2018chimp}, if they have visual impairments~\cite{zyskowski2015accessible,vashistha2018bspeak,randhawa2021karamad}, or if they come from a low-resource background~\cite{alsayasneh2017personalized,randhawa2021karamad}.



Two important aspects of collecting relevance judgements are: \textit{(i)} the quality of assessments~\cite{scholer2013effect}; and \textit{(ii)} the time taken by assessors to make their judgements~\cite{smucker2012time}. Since relevance judgements are used to train and evaluate \textit{Learning to Rank (LtR)} systems, the quality of judgements impacts the effectiveness of such systems~\cite{xu2010improving,clough2013evaluating}. The time taken by assessors to judge relevance may not only affect the quality of judgements, but also contribute to the cost of building (and maintaining) test collections. NIST assessors~\cite{craswell2020overview,craswell2021overview} and crowdworkers~\cite{alonso2009can,kutlu2020annotator} are often paid by their time spent on a task (e.g., as on \textit{Prolific}). The longer it takes assessors to judge, the costlier it becomes. 
There are a number of factors---not limited to topic difficulty~\cite{scholer2013effect,damessie2016influence}, document familiarity~\cite{scholer2011quantifying}, or relevance judgement session length~\cite{scholer2011quantifying}---that have been shown to affect the quality of (and the time taken for) judging relevance. 

In our work, we focus on two such factors in our pursuit to examine the feasibility of using the voice modality for text-document relevance assessments: \emph{document length}~\cite{saracevic1969comparative,singhal1996document,scholer2011quantifying,fu2022evaluating} and an assessor's \emph{cognitive abilities}~\cite{saracevic2007relevance,scholer2013effect} expressed in terms of working memory and inhibition. 
Our selection of factors is motivated by a range of prior works. The serial~\cite{lai2006speech} and temporal~\cite{schalkwyk2010your} nature of the voice medium makes it more difficult for listeners to \textit{``skim''} back and forth over a piece of information as compared to reading it on-screen~\cite{yankelovich1998designing,murad2020designing,xu2019waveear}. Voice interfaces also demand greater cognitive load when compared to text interfaces for processing information~\cite{shneiderman2000limits,lai2006speech,rajput2012evaluation}. These are exacerbated as the amount of information to be conveyed increases in size~\cite{nowacki2020improving,sherwani2007voicepedia}. 
Understanding how these factors affect the relevance judgement process can help us design tasks for assessors with a wide range of abilities and for different document presentation modalities. 
While there exists various measures for cognitive abilities, we selected two---\emph{working memory} (someone's ability to hold information in short-term memory)~\cite{diamond2013executive} and \emph{inhibition} (someone's ability to ignore or inhibit attention to stimuli that are not relevant)~\cite{diamond2013executive}---which have been shown to play an important role in speech understanding~\cite{rudner2012working,gordon2016effects,stenback2016speech}. We posit that they will also be crucial in the relevance judgement process, especially when documents are presented in the voice modality. 
Taken together, we investigate the following research questions.

\begin{itemize}[topsep=3pt]
    \item[\textbf{\rqOne{}}]{\textit{How does the modality of document presentation (text vs. voice) affect an assessor's relevance judgement in terms of accuracy, time taken, and perceived workload?}}
    \item[\textbf{\rqTwo{}}]{\textit{How does the length of documents affect assessors' ability to judge relevance?} Specifically, we look into the main effect of document length and the effect of its interplay with  presentation modality.}
    \item[\textbf{\rqThree{}}]{\textit{How do the cognitive abilities of an assessor (with respect to their working memory and inhibition) affect their ability to judge relevance?} Specifically, we look into the main effect of the cognitive abilities and the effect of their interplay with the presentation modality.}
\end{itemize}

       


     
To answer these questions, we conducted a quantitative user study ($n=49$) on the crowdsourcing platform Prolific. Participants judged the relevance of $40$ short and long documents sampled from the passage retrieval task data of the 2019 \& 2020 \textit{TREC Deep Learning (DL) track}~\cite{craswell2020overview,craswell2021overview}. Our findings are summarised as follows. 
\begin{itemize}[leftmargin=*,topsep=3pt]
\item Participants judging documents presented in the voice modality were \textit{equally} accurate as those judging them in the text modality. 
\item As documents got longer, participants judging documents in voice modality took significantly longer than those in text modality. For documents of length greater than $120$ words, the former took twice as much time with less reliable judgements.
\item  We also found that inhibition---or a participant's individual ability to ignore or inhibit attention to stimuli that are not relevant---impacts relevance judgements in voice modality. Indeed, those with higher inhibition were significantly more accurate than their lower inhibition counterparts.
\end{itemize}
Overall, our results indicate that we \emph{can} leverage the voice modality to effectively collect relevance labels from crowdworkers.

 \section{Related Work}
\subsection{Relevance Judgement Collection}

The general approach for gathering relevance
assessments for large document corpora (large enough that a full judgement of all corpus documents is not possible) was established by TREC in the early 1990s~\cite{harman1993overview}.
Given a set of information needs, a pooled set of documents based on the top-$k$ results of (ideally) a wide range of retrieval runs are assessed by topic experts. This method is typically costly and does not scale up~\cite{alonso2009can} once the number of information needs or $k$ increases. In the last decade, creating test collections using crowdsourcing via platforms like Prolific or \textit{Amazon Mechanical Turk (AMT)} have been shown to be a less costly yet reliable alternative~\cite{alonso2009can,scholer2011quantifying,zuccon2013crowdsourcing,kutlu2020annotator}. 
While the potential of crowdsourcing for more efficient relevance assessment has been acknowledged, concerns have been raised regarding its quality---as workers might be too inexperienced, lack the necessary topical expertise, or be paid an insufficient salary. In turn, these issues may lead them to completing the tasks to a low standard~\cite{kazai2013analysis,peer2017beyond,maddalena2017crowdsourcing}. Aggregation methods (e.g., majority voting) can be used as effective countermeasures to improve the reliability of judgements~\cite{jung2011improving,hosseini2012aggregating}. 

There are a number of factors that have been shown to affect the relevance judgement process. \citet{scholer2013effect} observed that participants exposed to non-relevant documents at the start of a judgement session assigned higher overall relevance scores to documents than when compared to those exposed to relevant documents.~\citet{damessie2016influence} found that for easier topics, assessors processed documents more quickly, and spent less time overall. Document length was also shown to be an important factor for judgement reliability.~\citet{hagerty1967abstracts} found that the precision and recall of abstracts judged increased as the abstract lengths increased (30, 60, and 300 words). In a similar vein,~\citet{singhal1996document} observed that the likelihood of a document being judged relevant by an assessor increased with the document length.~\citet{chandar2013document} found that shorter documents that are easier to understand provoked higher disagreement, and that there was a weak relationship between document length and disagreement between
the assessor. In terms of time spent for relevance judgement,~\citet{konstan1997grouplens} and~\citet{shinoda2012information} asserted that there is no significant correlation between
time and document length. On the other hand,~\citet{smucker2012overview} found participants took more time to read, as document length increased (from $\sim$10s for $100$ words, to $\sim$25s for $1000$ words). 

\subsection{Voice Modality}
Voice-based crowdsourcing has been shown to be more accessible for people with visual impairments~\cite{zyskowski2015accessible,vashistha2018bspeak}, or those from low resource backgrounds~\cite{randhawa2021karamad}. It can also provide greater flexibility to crowdworkers by allowing them to work in brief sessions, enabling multitasking, reducing effort required to initiate tasks, and being reliable~\cite{vashistha2017respeak,hettiachchi2020hi}. However, information processing via voice is inherently different compared to when it is presented as text. The use of voice has been often shown to lead to a higher cognitive load~\cite{murad2018design,vtyurina2020mixed}. Individuals also exhibit different preferences. For example,~\citet{trippas2015towards} observed that participants preferred longer summaries for text presentation. For voice however, shortened summaries were preferred when the queries were single-faceted. Although their study did not measure the accuracy of judgements against a ground truth, what participants considered the most relevant was similar across both conditions (text vs. voice presentation). Furthermore, the voice modality can leverage its own unique characteristics for information presentation. For instance,~\citet{chuklin2019using} varied the prosody features (pauses, speech rate, pitch) of sentences
containing answers to factoid questions. They found that emphasising the answer phrase with a lower speaking rate and higher pitch increased the perceived level of information conveyed.

Concerning the collection of relevance assessments,
~\citet{tombros1999study} found in their lab study that participants were more accurate and faster in judging relevance when the list of documents (with respect to a query) were presented as text on screen as compared to when they were read out to the participants---either in person, or via telephone. It should however be noted that this work was conducted more than two decades ago---barely ten years after the invention of the Web, when the now common voice assistants and voice-enabled devices were long to be developed.

The work closest to ours is the study by~\citet{vtyurina2020mixed}, who presented crowdworkers with five results of different ranks from \textit{Google}---either in text or voice modality. They asked their participants to select the two most useful results and the least useful one. They observed that the relevance judgements of participants in the text condition were significantly more consistent with the true ranking of the results than those who were presented with five audio snippets. The ability to identify the most relevant result was however \textit{not} different between the experimental cohorts. This study did not consider the effect of document length or cognitive abilities of participants on their relevance judgement performance, which is what we explore.

\subsection{Cognitive Abilities}

 
Prior works have explored how the cognitive abilities of assessors impact relevance judgements.~\citet{davidson1977effect} observed that openness to information---measured by a number of cognitive style variables such as open-mindedness, rigidity, and locus of control---accounted for approximately 30\% of the variance in relevance assessments.~\citet{scholer2013effect} found that assessors with a higher need for cognition (i.e., a predisposition to enjoy cognitively demanding activities) had higher agreement with \textit{expert} assessors, and took longer to judge compared to their lower need for cognition counterparts. Our work focuses on \textit{working memory} and \textit{inhibition}.

\vspace*{1mm}\textbf{\textit{Working Memory (\WM{})}}
 refers to \textit{an individual's capacity for keeping information in short-term memory even when it is no longer perceptually present}~\cite{diamond2013executive}. This ability plays a role in higher-level tasks, such as reading comprehension~\cite{lustig2001working} 
 and problem solving~\cite{wiley2012working}. \citet{macfarlane2012phonological} observed that participants with dyslexia---a learning disorder characterised by low working memory---judged fewer text documents as non-relevant when compared to participants without the learning disorder. They posited that it might be cognitively more demanding to identify text documents as non-relevant for the cohort with dyslexia.
 With regards to processing speech, High \WM{} has also been shown to be helpful in adapting to distortion of speech signals caused by background noise~\cite{gordon2016effects}.~\citet{rudner2012working} and~\citet{stenback2016speech} observed high \WM{} individuals perceived less effort while recognising speech from noise. 

\vspace*{1mm}\textbf{\textit{Inhibition (\In{})}} refers to the capacity to regulate attention, behaviour, thoughts, and/or emotions by overriding internal impulses or external \textit{`lure'}---and maintaining focus on what is appropriate or needed~\cite{diamond2013executive}. To our knowledge, prior studies have not investigated the effect of \In{} on the relevance assessment process. High \In{} has been shown to help in speech recognition, especially in adverse conditions like the presence of background noise~\cite{stenback2016speech,stenback2021contribution}. 
\vspace{1mm}

A significant number of prior works have explored various aspects related to the process of relevance assessment. This work however considers the novel effect of document length and the cognitive abilities of assessors to explore the utility of the voice modality with regards to judging relevance.

 \section{Methodology}\label{sec:method}
To address our three research questions outlined in \S\ref{sec:intro}, we conducted a crowdsourced user study. The study participants were asked to judge the relevance of \textit{Query/Passage (Q/P)} pairings, where passages were presented either in the form of \textans{} (i.e., a piece of text) or \voiceans{} (i.e., an audio clip). In our study, passage \textbf{presentation modality} is a \textit{between-subjects} variable. We also controlled the \textbf{length of passages}; this is a \textit{within-subjects variable} to ensure that participants judged passages of varying lengths. The \emph{independent variables} \textbf{working memory} and \textbf{inhibition} allow us to estimate the impact of the cognitive abilities of the participants on the accuracy of their judgements, time taken and perceived workload.

\begin{figure}[!t]
    \includegraphics[width=\linewidth]{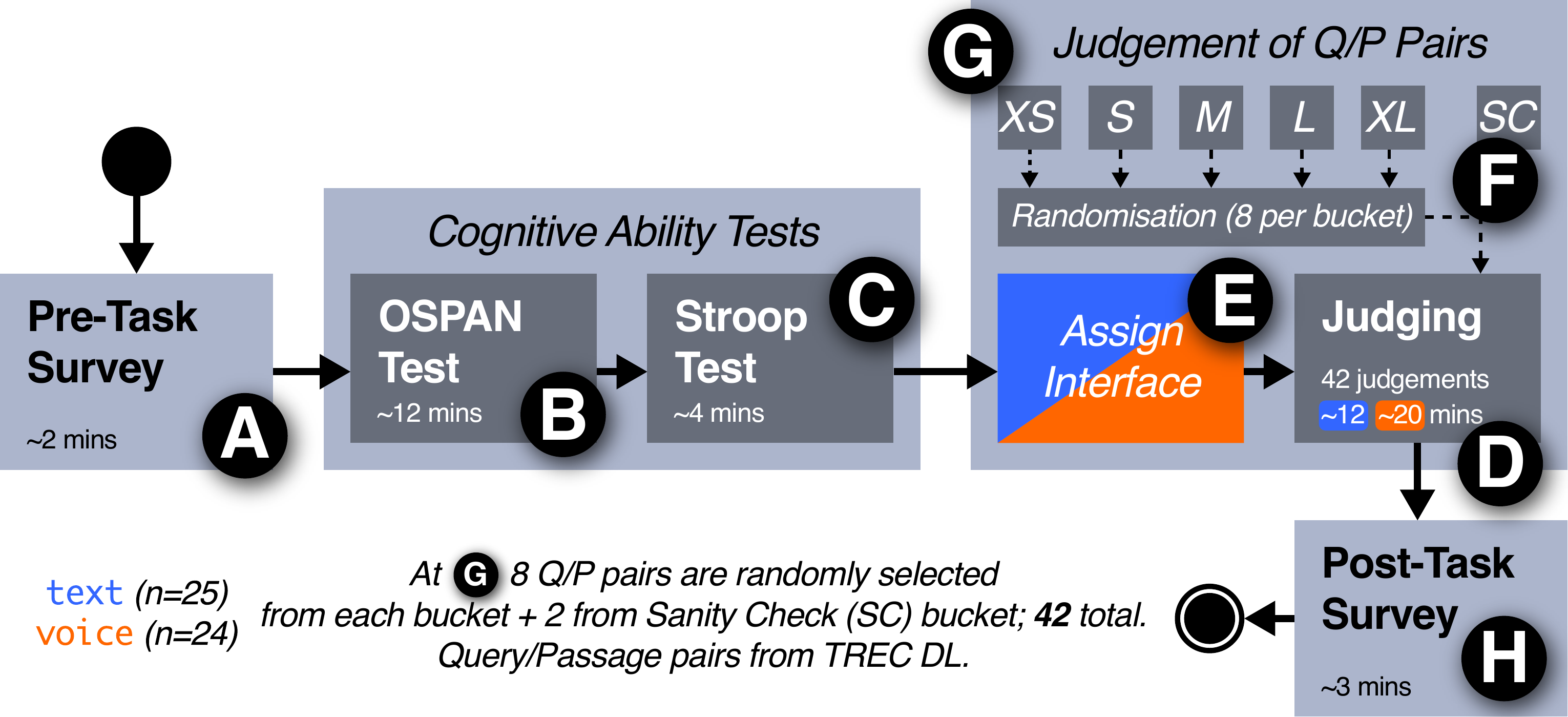}
    \caption{A high-level overview of the user study protocol, including approximate times for participants to complete each component. Refer to \S\ref{sec:exp_proc} for mappings to the letters highlighting key aspects of the study procedure.}
    \label{fig:study}
     \vspace{-3mm}
\end{figure}


\subsection{Study Overview}\label{sec:exp_proc}

Figure~\ref{fig:study} presents an overview of the user study design.\footnote{Note that circles refer to superimposed labels on the illustration in Figure~\ref{fig:study}.} The diagram highlights the main tasks that study participants undertook. Lasting approximately 32 minutes for \textans{} and 40 minutes for \voiceans{}, the study consisted of four main parts: \textit{(i)} the \textit{pre-task survey} (\S\ref{sec:demographics}); \textit{(ii)} the \textit{cognitive ability tests} (\S\ref{sec:catests}); \textit{(iii)} the \textit{judgements} (\S\ref{sec:interface}); and \textit{(iv)} the \textit{post-task survey} (\S\ref{sec:measures}).

After agreeing to the terms of the study, participants completed a pre-task survey \circled{A}. This survey included demographics questions, including questions about their familiarity with voice assistants---as reported in \S\ref{sec:demographics}. Participants would then move onto two \textit{psychometric tests}; as outlined in \S\ref{sec:catests}, these tests measured their cognitive abilities with respect to working memory \circled{B} and inhibition \circled{C}. Participants undertook a short practice task to help them familiarise themselves with the interface for each test.

After the psychometric tests, participants moved to the main part of the study: judging Q/P pairings \circled{D}. The experimental system first assigned the participants to either 
\textans{} or \voiceans{} randomly \circled{E} (\S\ref{sec:interface}). Based on the assigned condition, participants then judged a total of 42 {Q/P} pairings presented to them in a random order to mitigate the effect of topic ordering~\cite{scholer2011quantifying,scholer2013effect} (\S\ref{sec:qa})---$40$ were selected from the \textit{2019 and 2020 TREC Deep Learning (DL) track}, and the remaining two acted as a \textit{sanity check} (\SC{}) \circled{F}. The $40$ passages belonged to different \textit{answer length buckets} \S\ref{sec:qa} \circled{G}. Finally, the participants would be taken to the post-task survey \circled{H}.

\subsection{Query/Passage Pairings}\label{sec:qa}
As mentioned, we obtained the Q/P pairings from the 2019 and 2020 TREC DL track---specifically the passage retrieval task~\cite{craswell2020overview,craswell2021overview,mackie2021deep}. The test partition of the datasets contain 43 and 54  natural language queries with passages that are judged by \textit{NIST} assessors. Using a graded relevance scale, passages for each query were judged by assessors as: \textit{(i)} \textit{perfectly relevant} when the passage is dedicated to the query, containing an exact answer; \textit{(ii)} \textit{related} when  the passage appears somewhat related to the query, but does not answer it completely; or \textit{(iii)} \textit{non-relevant}, when the passage has nothing to do with the provided query~\cite{craswell2020overview,craswell2021overview}. We note that an additional relevance category exists (\emph{highly relevant}). However, we ignore judgements of this category in our work (similar to~\cite{kutlu2020annotator}) in order to have a clear distinction between the different categories.

\paragraph{\textbf{Sampling Procedure}}
From the available test queries, we sampled $40$ (due to budget constraints). As \textbf{RQ2} states we are interested in how passage length affects assessments, we next determined five different buckets of passage length: from \emph{very short} to \emph{very long} (more details follow below). We randomly assigned the 40 queries to these five buckets, leading to eight queries per passage length bucket. For each query, we sampled three passages from the QRELs, with the additional condition that the sampled passages must fall into the query's passage length bucket: one \emph{perfectly relevant}, one \emph{related}, and one \emph{non-relevant} passage. And thus, each bucket contains $24$ passages pertaining to eight queries. Table~\ref{tab:info} demonstrates three Q/P examples, each coming from a different passage length bucket.

\paragraph{\textbf{Sanity Check (\SC{})}} We also created two additional Q/P pairings to act as a sanity checks\footnote{The sanity check questions were: \textit{(i)} \textit{Who was the lead vocalist of Queen?}, with the answer passage being perfectly relevant; and \textit{(ii)} \textit{What is the difference between powerlifting and weightlifting?}, with the answer passage being non-relevant.} in order to perform quality control of the relevance judgements by our participants, as suggested by~\citet{scholer2011quantifying}. \textit{We did not consider the \SC{} Q/P pairs in our data 
analysis.}

\paragraph{\textbf{Judgements per Participant}}
We presented all our participants with the \textit{same} set of 40 queries + 2 \SC{} queries in order to mitigate effects arising due to differences in queries~\cite{damessie2016influence}. Each participant judged one randomly sampled passage---out of the three available ones---for each of the 40 queries (ignoring the \SC{} queries). We thus collected relevance judgements on a total of $40\times{}3 = 120$ Q/P pairs\footnote{The list of collected Q/P pairs are available \href{https://osf.io/48vx5/?view_only=9ed09286e3b74c6c853e24411b798826}{here}.}. Each participant judged ~13 passages per QREL.

\paragraph{\textbf{Passage Length Buckets}} To add more detail to our passage length bucketing procedure, we chose five types of length buckets: \XSh{} \textit{(Very Short)}; \Sh{} \textit{(Small)}; \M{} \textit{(Medium)}; \Lo{} \textit{(Long)}; and \XLo{} \textit{(Very Long)}. They corresponded to the $0-5$, $5-50$, $50-75$, $75-99$ and $99+\%$-ile of the lengths of all judged passages of the $97$ test queries in our TREC-DL datasets. We selected the percentiles to have a range of $20$ to $30$ words per passage length bucket. The concrete word ranges for each passage length bucket can be found in Table~\ref{tab:ans}.

 \begin{table}[t!]
     \centering
    \footnotesize
     \caption{Overview of passage length buckets. Averages are reported together with the standard deviation. 
     }
     \label{tab:ans}
    
     \renewcommand{\arraystretch}{1.25}
    
    \begin{tabularx}{\linewidth}{X||Y|Y|Y|Y}

          \textbf{{Passage} Length}& 
          \textbf{Min-max \#words}& 
         \textbf{Avg. \#words} & \textbf{Min-max audio clip length (s)} &
         \textbf{Avg. audio clip length (s)}    \\ \hline\hline
          \textbf{Very Short}  &   $ 12 - 32 $ &  $24.67 (\pm5.3)$ &  $ \ \ 3 - 13 $ &  $ 10.04 (\pm 2.4 )$ \\ \hline
        \textbf{Short} &  $ 33 - 53 $ &  $41.67 (\pm 3.8)$& $ 14 - 19 $&  $ 17.04 (\pm 1.4 )$ \\ \hline          \textbf{Medium} &    $ 54 - 74 $ &  $63.17 (\pm 5.6)$ &  $ 20 - 30 $ &  $ 25.17 (\pm 3.4 )$ \\ \hline
         \textbf{Long} & $ 90 - 120 $ & 
         $99.79 (\pm 6.9) $& $ 31 - 42 $ & $ 36.04 (\pm 3.04 )$ \\  \hline
          \textbf{Very Long}  &    $ 121 - 151 $ &  $139.96 (\pm 8.2) $ &  $48 - 70 $ &  $ 54.58 (\pm 4.9 )$\\ 
         
    \end{tabularx}

    \vspace*{-3mm}
\end{table}
\begin{table*}[t!]
    \centering
    \caption{Examples of \textit{Query/Passage (Q/P)} pairs for different passage length categories. The (\texttt{Qid}) is taken from the TREC datasets. We also provide links to [audio \faIcon[regular]{file-audio}] clips of the respective passages. 
    }
    \vspace*{-2mm}
    \label{tab:info}
    
    \renewcommand{\arraystretch}{1.2}
    \footnotesize
    \begin{tabular}{>{\raggedright}p{1.7cm}||>{\raggedright}p{2.9cm}>{\raggedright}p{11.9cm}}
        \textbf{Passage Length} &
        \textbf{Query (\texttt{Qid})} & \textbf{ \colorbox{lightgray}{Ground Truth Relevance}  \& Passage} 
 \tabularnewline
        \hline\hline
        \multirow{2}{=}{{\textbf{Very Short (\XSh{})}}} & {{\textit{What metal are hip replacements made of?} (\texttt{877809}) }} & \colorbox{lightgray}{\texttt{RELEVANT}}
        Some prosthesis, like hip and knee joints made of cobalt chrome, contain some trace of nickel and for patients with allergies to this may have to go with Titanium joints. [\href{https://drive.google.com/file/d/1sckq16K78A_wY8QP74E-e3CvnPE3SpoC/view?usp=sharing}{Audio \faIcon[regular]{file-audio}}] \tabularnewline \hline
       
        \multirow{3}{=}{{\textbf{Short (\Sh{})}}} & \textit{Who has the highest career passer rating in the nfl?} (\texttt{1056416})&\colorbox{lightgray}{\texttt{SOMEWHAT-RELEVANT}} Wilson is the only quarterback in NFL history to post a 100-plus passer rating in each of his first two seasons, and he's already won a Super Bowl. Dan Marino is really the only quarterback you could argue was better out of the gate. [\href{https://drive.google.com/file/d/1VCAsBzh6M_KJwiC8YD488dge1tBq-cTC/view?usp=sharing}{Audio \faIcon[regular]{file-audio}}]  \tabularnewline \hline

       
        \multirow{4}{=}{{\textbf{Long (\Lo{})}}} & \multirow{4}{=}{{\textit{What is the appearance of granulation tissue?} (\texttt{1133579})}} & \colorbox{lightgray}{\texttt{NON-RELEVANT}} The protective outer layer of the plant. Everything needs skin, or at least some sort of a covering, for plants, it's a system of dermal tissue. Which covers the outside of a plant and it protects the plant in a variety of ways. Dermal tissue called epidermis is made up of live parenchyma cells in the non-woody parts of plants. Epidermal cells can secrete a wax-coated substance on leaves and stems, which becomes the cuticle. Dermal tissue that is made up of dead parenchyma cells is what makes up the outer bark in woody plants. [\href{https://drive.google.com/file/d/1vySeF5ob6juBgENn-apcJzK8WJCsA5aP/view?usp=sharing}{Audio \faIcon[regular]{file-audio}}] 
        \tabularnewline

    \end{tabular}
    \vspace*{-3mm}
\end{table*}

\paragraph{\textbf{From Text Passage to Audio Clip}} We processed the passages to remove any unwanted punctuation, leading and trailing whitespace, and corrected a few spelling errors. These cleaning steps were necessary as we did not want the participants to be distracted by unclean text, and to create legible audio clips for the \voiceans{} interface. We used \textit{Amazon Polly}\footnote{https://aws.amazon.com/polly/}---an open-source text to speech system with an array of options for language and voice types---to generate the audio clips for the voice results. Specifically, we chose \texttt{Matthew}, a male US English voice, with a speed of 95\% as the authors unanimously agreed that this particular setting (among other evaluated voice options)  had the clearest pronunciation, in particular of difficult words\footnote{Difficult words in this context include words from languages other than English (e.g., \textit{``..and include Gruy\`{e}re, Emmental, T\^{e}te De Moine, Sbrinz..''}), words specific to a domain (e.g., \textit{``...the manubrium, sternebrae, and xiphoid cartilage.''}), etc.} that might appear in the passages. Lastly, we ran a pilot study ($n=5$) where participants were asked to rate the pace, accent, and length of our generated audio clips on a seven-point scale. They reported an average score of $6.3$, confirming the high quality of the audio clips for our task. Table~\ref{tab:ans} shows the minimum, maximum, and average length of the audio clips in seconds for the passages belonging to the five length buckets.\footnote{Audio clips for all the passages are released \href{https://osf.io/48vx5/?view_only=9ed09286e3b74c6c853e24411b798826}{here}.}

\subsection{Cognitive Ability Tests}\label{sec:catests}
In order to measure the cognitive abilities of our participants with relation to judging the presented Q/P pairings, we chose two established psychometric tests that examine both an individual's working memory and their inhibition. Prior work~\cite{rudner2012working,gordon2016effects,stenback2016speech} has shown that working memory and inhibition play an important role in speech understanding. 

\paragraph{\textbf{Working Memory}}
To measure working memory capacity, we used the \textit{Operation-word-SPAN (OSPAN)} test~\cite{turner1989working} that has also been used in prior \textit{Interactive IR (IIR)} work~\cite{choi2019effects}. The OSPAN test measures an individual's ability to recall letters displayed in sequence, while concurrently completing simple secondary tasks. Participants completed eight trials of varying lengths. During each trial, participants were shown a sequence of $3-7$ letters, and were then asked to recall the letters in their original order from a grid display. Additionally, during each trial, participants completed simple mathematical problems between each letter shown in sequence (e.g., \textit{``is 8+6=15?''}). The final score was equal to the sum of sequence lengths of all trials perfectly recalled. A higher score in the OSPAN test indicates a participant's greater ability to hold information (the letter sequence in correct order) in short-term memory when it is no longer perceptually present.

\paragraph{\textbf{Inhibition}} 
To measure inhibition, we used the \textit{Stroop test} which was first introduced in 1935~\cite{stroop1935studies}. As an example, the Stroop test has been used to measure inhibitory attention control in learning~\cite{kane2003working,gass2013inhibitory} and speech processing~\cite{stenback2016speech}. We used a computerised version of the test that was also used in the IIR study undertaken by~\citet{arguello2019effects}.
During the Stroop test, participants were shown a sequence of words indicating
one of four colours: red, green, yellow, or blue. Some of the words
displayed are congruent (e.g., the word \texttt{``\textcolor{blue}{blue}''} displayed in blue font), and others are incongruent
(e.g., the word \texttt{``\textcolor{red}{blue}''} displayed in red font). For each word, participants had to indicate the \textit{font colour} of the word as quickly as possible by clicking on the correct option presented as a list (the trial continued until the correct colour was chosen). Participants had to complete $48$ correct trials (similar to the study by~\citet{arguello2019effects}), of which $24$ are congruent and $24$ are incongruent. The final score is equal to the participant's average response time (in milliseconds) for the incongruent trials, minus the average response time for the congruent trials. Response times are typically slower for the incongruent trials, an effect referred to as the \textit{Stroop effect}. Lower scores are better for the Stroop test, with higher scores indicating a greater difficulty in focusing on the relevant stimulus (the colour of the word) and ignoring the non-relevant stimulus (the word itself).

\begin{figure*}[!t]
    \includegraphics[width=\linewidth]{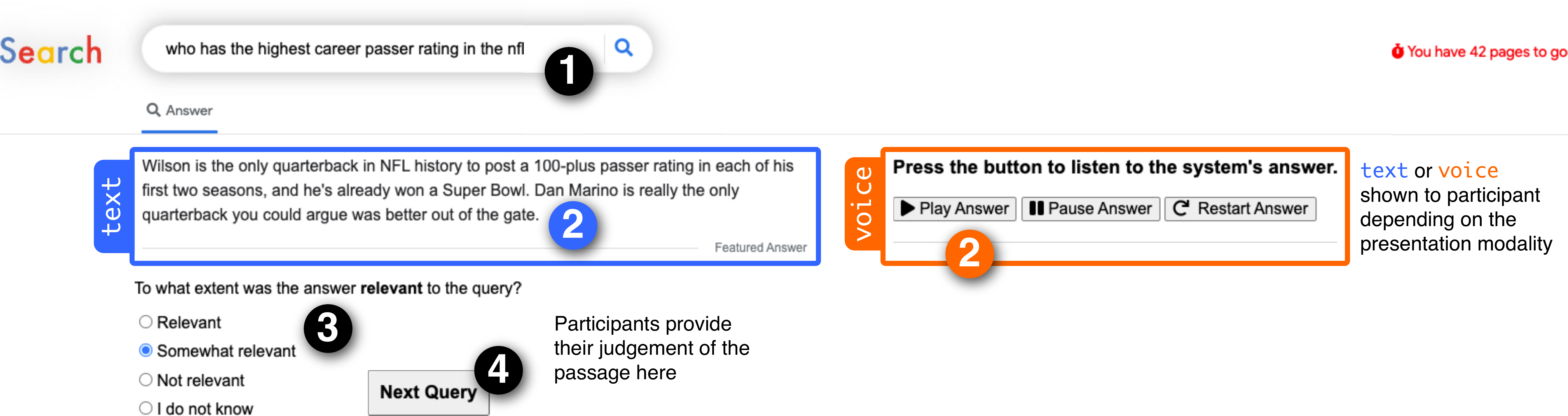}
    \vspace*{-4mm}
    \caption{Composition screenshot of both the {\color{textcolour}{\texttt{text}}} and {\color{voicecolour}{\texttt{voice}}} interfaces used by participants for judging query-passage pairs. Circled numbers correspond to the same in the narrative, found in \S\ref{sec:interface}.}
    \vspace*{-3mm}
    \label{fig:rel_temp}
\end{figure*}

\subsection{Assessor Interface}\label{sec:interface}
Our study interface is shown in Figure~\ref{fig:rel_temp}, as a composition of both the \textans{} and \voiceans{} interfaces. The \textans{}-specific components are highlighted in {\textcolor{textcolour}{blue}}; \voiceans{}-specific ones in {\textcolor{voicecolour}{orange}}. For each Q/P pairing they were required to judge, participants were presented with a static query box~\circled{1} which could not be altered; it displayed the query for which the participant was to judge the passage for. Only one passage was shown~\circled{2}; depending on the condition, this was either presented as text (for \textans{}), or a series of buttons to control the audio clip (for \voiceans{}). 
In the case of \voiceans{}, the participant had to press the \texttt{Play Answer} button to listen to the audio clip. They could also pause and restart the audio clip by pressing the \texttt{Pause Answer} and \texttt{Restart Answer} buttons respectively.

Once they had read or listened to the answer passage, participants then moved to the underlying form located at~\circled{3} to provide their judgement of the passage. Participants could choose between \textit{`Relevant'}, \textit{`Somewhat relevant'}, \textit{`Non relevant'}, and \textit{`I do not know'}. We included the final option to ensure that participants were not forced to make a relevance decision in the case that they were not sure as it has been shown that assessors are not always certain of their judgements~\cite{al2014qualitative}. We did not provide the participants with the option to skip parts of the audio clip or adjust the speed. Certain checks were in place to ensure reliability of relevance judgements of participants, \textit{in addition} to the two \SC{} pairings as outlined in \S\ref{sec:qa}. For \textans{}, the form for marking relevance~\circled{3} appeared after five seconds. For \voiceans{}, the form for marking relevance~\circled{3} appeared after 50\% of the audio clip had been played. Participants could also proceed to judge the next query/passage pair by clicking the \textit{Next Query} button~\circled{4} which was enabled only after a participant made their judgement. Once participants moved on to the next pairing, they could not go back to revise earlier judgements. No time limit was imposed on participants during the judging process.

\subsection{\textbf{Outcome Measures}}\label{sec:measures}
In addition to the use of the two psychometric tests outlined in \S\ref{sec:catests}, we used interaction logging apparatus and additional surveys to capture both behavioural and experience data respectively.

\paragraph{\textbf{Measuring Participant Behaviours}}
We added the \textit{JavaScript} library \logui{}~\cite{maxwell2021logui} into our web-based judgement interface; it allowed us to capture a variety of different behaviours and events such as:
\textit{(i)} when the page was loaded; \textit{(ii)} clicks on the form to record the judgement made by a participant; and \textit{(iii)} clicks on the \texttt{Play}/\texttt{Pause}/\texttt{Restart} buttons (for \voiceans{}). From these events, we could compute the amount of time taken for an individual to make a judgement---that is, from when the page loaded (showing the query/passage pairing) to when the \texttt{Next Query} button was clicked \circled{4} (Figure~\ref{fig:rel_temp}). In turn, this allowed us to compute the \textit{time per relevance judgement}, as reported in our results.

\paragraph{\textbf{Measuring Participant Experiences}}
After completing the relevance judgements, participants completed the post-task survey. Participants were asked about their perceived workload. They were asked specifically to answer the questions based \textit{only} on their perceived experiences of the relevance judgement tasks. To measure workload, we used five questions from the raw \textit{NASA TLX} survey, as proposed by~\citet{hart1988development}.  This instrument has been used (in slightly different forms) in several prior IIR studies (e.g.,~\cite{arguello2012task,arguello2019effects,roy2022users}). The five selected questions from the NASA TLX are designed to measure perceived: \textit{(i)} mental demand; \textit{(ii)} effort; \textit{(iii)} temporal demand; \textit{(iv)} frustration; and \textit{(iv)} performance. We omitted the \textit{`physical demand'} question from the survey as it was not relevant to our task.\footnote{This was also done in prior studies, such as the study reported by~\citet{vtyurina2020mixed}} Participants responded to the five NASA TLX questions using a seven-point scale with labelled endpoints (from \textit{``poor''} to \textit{``good''} for performance and from \textit{``low''} to \textit{``high''} for the remaining four). 


\paragraph{\textbf{Measuring Participant Performance}}
We also computed the \textit{accuracy} of our participants in the relevance judgement tasks. Accuracy 
was calculated in terms of how many Q/P pairs participants judged \textit{correctly}---that is, their relevance judgement matching the ground truth from the QRELs. 
We also aggregated relevance judgements of participants on each Q/P pairing based on majority voting, as done by~\citet{kutlu2020annotator} to observe if collective judgements are more accurate. We used Krippendorff's alpha ($\alpha$) to measure inter-annotator agreement (as used by~\citet{damessie2016influence}). Lastly, we calculated Cohen's kappa ($\kappa$) ~\cite{carletta1996assessing,artstein2008inter,bailey2008relevance} which measures the agreement of judgements with ground truths by considering chance. 

\subsection{Participant Demographics}\label{sec:demographics}

We conducted an \textit{a-priori} power analysis using \textit{G-power}~\cite{faul2007g} to determine the minimum sample size required to test our \textbf{RQ}s. The results indicated that the required sample size---to achieve $95\%$ power for detecting an effect of $0.25$, with two groups (modality) and five measurements (passage length)---is $46$.
As such, we recruited $50$ participants from the Prolific platform. We disqualified one participant as they failed to correctly judge our sanity check query/passage pairs (\S\ref{sec:qa}). Our $n=49$ (25 for \textans{}{}, 24 for \voiceans{}) participants were native English speakers, with a $98\%$ approval rate on the platform---a minimum of $250$ prior successful task submissions, and self-declared as having no issues in seeing colour. Participants were required to use a desktop/laptop device in order to control for variables that might affect results of the Stroop and OSPAN tests on other (smaller) devices. From our participants, $22$ identified as female, $24$ as male, with $3$ declining to disclose this information. The mean age of our participants was $38$ (\textit{min.} $22$, \textit{max.} $69$). With respect to the highest completed education level, $28$ possessed a Bachelors (or equivalent), nine has a Masters (or equivalent), ten had a high school degree, and two had a PhD (or equivalent). We also asked participants how often they used a smart speaker to search for information, and listening to the provided answer---to which $13$ reported daily usage, $20$ reported usage on a weekly basis, and $16$ said never. 
Participants were paid at the rate of GBP\textsterling{}11/hour, a value that is greater than the 2022-2023 \textit{[outside London] UK Real Living Wage}.

 \section{Results and Discussion}
This section presents the results of our experiments pertaining to our three \textbf{RQ}s. First, we provide details on the statistical tests we conducted, and how we utilised the cognitive ability tests to divide participants into \textit{low-} and \textit{high-ability} groups.

\paragraph{Statistical Tests} 
For our analyses\footnote{All data and code pertaining to our analyses are \href{https://osf.io/48vx5/?view_only=9ed09286e3b74c6c853e24411b798826}{released}.}, we conducted a series of independent sample \textit{t}-tests with Bonferroni correction ($\alpha=0.05$) to observe if the modality of presentation has a significant effect on our dependent variables---accuracy of relevance judgements, the time taken to judge, and the perceived workload (\textbf{RQ1}). We also conducted a series of mixed factorial ANOVA tests (where modality of presentation is a \textbf{\textit{between-subjects}} variable, and passage length is a \textbf{\textit{within subjects}} variable) to observe if presentation modality, passage length, or the interaction between them have a significant effect on accuracy of relevance judgement and time taken (\textbf{RQ2}). Lastly, we conducted a series of three-way ANOVA tests to observe if the two user dispositions---working memory and inhibition--or their interaction with modality of presentation have a significant effect on the three dependent variables (\textbf{RQ3}). For \textbf{RQ2} and \textbf{RQ3}, we followed up the ANOVA with pairwise Tukey tests with Bonferroni correction ($\alpha=0.05$) to observe where significant differences lay. In the case where no significant difference was observed between the two conditions, we used equivalence testing between conditions through the \textit{two one-sided t-tests (TOST)} procedure. The upper and lower bounds for the TOST was set at $7.5\%$ (-$\Delta$L = $\Delta$U = 7.5) for accuracy, as~\citet{xu2010improving} observed that LtR models were robust to errors of up to $10\%$ in the dataset (we used $7.5\%$ for \textit{conservativeness}). For each scale of NASA-TLX, we set -$\Delta$L = $\Delta$U = 2.04, following~\citet{lee2021s}, who used a bound of $\pm 18$ on a $100$-point NASA TLX. For our seven-point scale, it translates to $\pm 2.08$ 
according to the formula of~\citet{hertzum2021reference}.

\paragraph{Cognitive Ability Scores and High vs. Low Ability Groups}
 To examine the effect of a participant's cognitive abilities on relevance judgement accuracy (\textbf{RQ3}), we performed a median split of the scores obtained by the participants in the OSPAN (\textit{min.} $0$, \textit{max.} $50$, \textit{mean} $=25.4(\pm 12)$, \textit{median} $=22$) and Stroop test (\textit{min.} $=-300$, \textit{max.} $=650$, \textit{mean} $=171.25(\pm 184)$, \textit{median} $=170$) respectively. The mean scores of our participants for working memory and inhibition were within one standard deviation of the reference mean scores as reported in~\cite{arguello2019effects}, validating our methodology. Participants were thus divided into a high- and low-ability group for each of working memory (based on OSPAN test scores) and inhibition (based on Stroop test scores). Note that for inhibition, a low test score indicates high ability. Prior studies have also analysed the effects of different cognitive abilities by dividing participants into low/high ability groups using a median split~\cite{al2011effect,scholer2013effect,arguello2019effects,choi2019effects}.

\subsection{RQ1: Modality of Passage Presentation}\label{sec:RQ1}

    \begin{table}[!t]
     \centering
    
     \caption{\textbf{RQ1:} Effect of modality of passage presentation on accuracy of relevance judgement, time taken per judgement in seconds and perceived workload (IV-VIII) per participant. We also report Krippendorff's $\alpha$ and Cohen's $\kappa$ for accuracy. $\dagger$ indicates significant difference in between the two conditions according to independent sample t-test.  $\star$ indicates the corresponding metric is equivalent for both conditions based on the TOST procedure. 
     }
     \vspace*{-2mm}
     \label{tab:RQ1}
    
     \renewcommand{\arraystretch}{1.4}
    \footnotesize
    \begin{tabularx}{\linewidth}{>{\hsize=0.1\hsize}Y>{\hsize=1.5\hsize}X||>{\hsize=1.2\hsize}Y|>{\hsize=1.2\hsize}Y}
         
          & \textbf{Metrics} & 
         \textans{} &
         \voiceans{}   \\
         \hline\hline
        \textbf{I} &   \textbf{Accuracy} $\star$ &  $68.40 (\pm 9.15)\%$ &  $65.94 (\pm 8.56)\%$ \\ 
      &    $\mathbf{\alpha}$\textbf{,} $\mathbf{\kappa}$ &  0.41, 0.61 &  0.37, 0.54 \\ \hline
\textbf{II} & \textbf{Majority Voting Acc.} & 79.1\% & 75.8\% \\ 
 &  $\mathbf{\kappa}$ & 0.76 & 0.71 \\ \hline 
 \textbf{III} &  \textbf{Time/Rel. Judge. (sec.)} $\dagger$  &  $17.56 (\pm 9.08)$ &  $29.54 (\pm 7.85)$ \\ \hline

  \textbf{IV}   &   \textbf{ Mental demand}$\star$ &  $4.68 (\pm 1.60)$ & $4.83 (\pm 1.37)$ \\ 
 \textbf{V} &  \textbf{Effort}$\star$&   $4.88 (\pm 1.88)$ &  $4.00 (\pm 1.50)$ \\ 
\textbf{VI} & \textbf{Temporal Demand}$\star$ &  $4.04 (\pm 1.86)$ & $3.08 (\pm 1.82)$ \\ 
  \textbf{VII} &   \textbf{Frustration}$\dagger$ &   $3.96 (\pm 2.07)$ &   $1.83 (\pm 0.82)$ \\ 
\textbf{VIII} & \textbf{Performance}$\dagger$ & $4.16 (\pm 1.93)$ & $5.67 (\pm 0.70)$ 
\\ 
    \end{tabularx}
    \vspace*{-5mm}
\end{table}

\begin{figure}
\centering

\includegraphics[width=\linewidth]{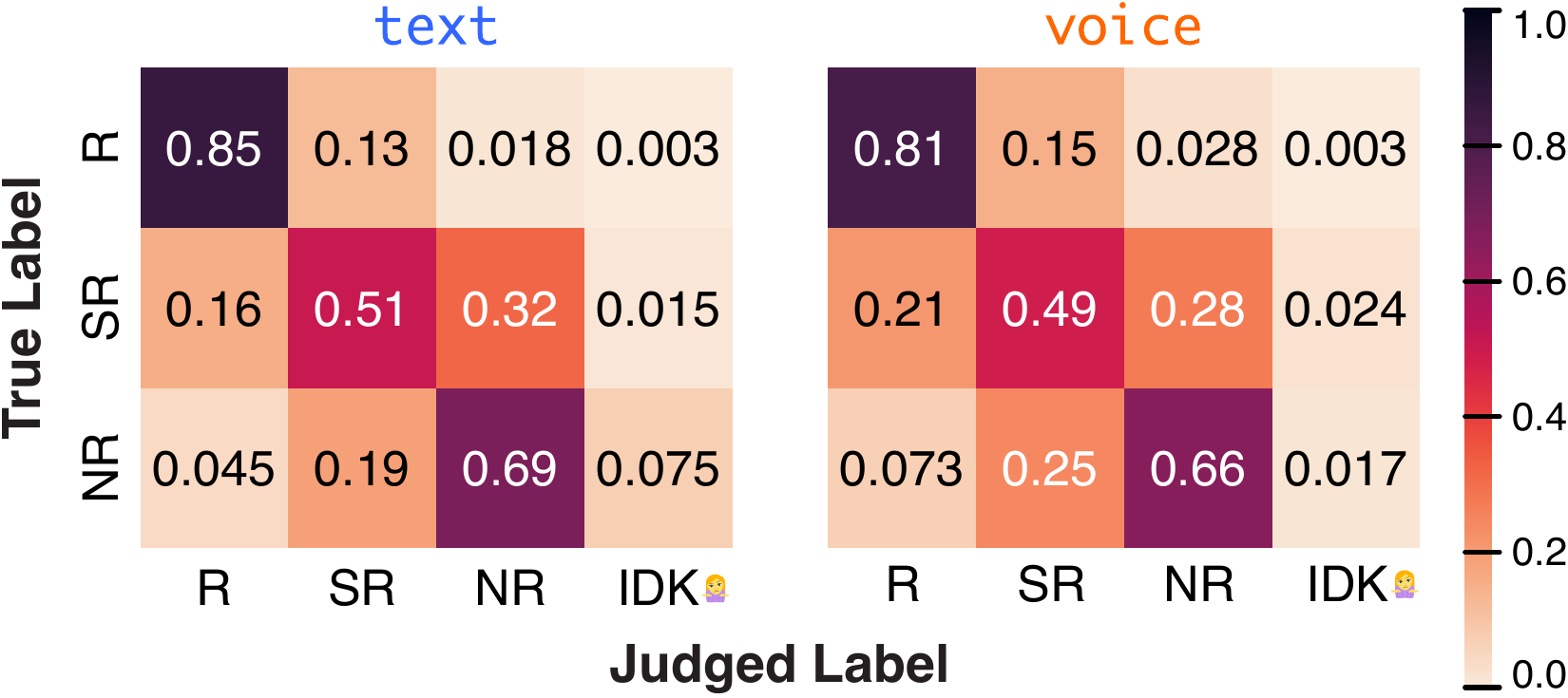}

%
%
\caption{Accuracy of relevance judgements per label category for both \textans{} and \voiceans{}. Diagonals represent percentage of time the true labels were \textit{correctly} predicted by participants.
Here, \texttt{R} = \texttt{RELEVANT}, \texttt{SR} = \texttt{SOMEWHAT-RELEVANT}, \texttt{NR} = \texttt{NON-RELEVANT} and \texttt{IDK} = \textit{I do not know}.}
\label{fig:relerrors}
\vspace{-0.5cm}
\end{figure}

Table~\ref{tab:RQ1} presents the main results for \textbf{RQ1}. There was no significant difference in judgement accuracy (row \textbf{I}, Table~\ref{tab:RQ1}) between participants in \textans{} and those in \voiceans{} ($t(47) = 0.97, p=0.33$). TOST revealed that accuracy of judgements across both conditions were \textit{equivalent} ($p=0.02$). The inter-annotator agreement ($\alpha$) was slightly higher in \textans{}. When using majority voting to aggregate relevance judgements (on average we had eight judgements per Q/P pair in each condition), we found that the accuracy increased from $68\%$ and $66\%$ to $79\%$ and $76\%$ respectively for \textans{} and \voiceans{} (\textbf{II}, Table~\ref{tab:RQ1}). This observation is in line with prior work~\cite{kutlu2020annotator}, which shows that aggregating judgements from several assessors is more reliable than a single untrained assessor. Cohen's $\kappa$ also increased with majority voting for both experimental conditions, indicating an increase in judgement reliability. Participants also showed similar trends of relevance judgement accuracy per relevance label category for both  experimental conditions. As shown in Figure~\ref{fig:relerrors}, participants in both conditions were most accurate in judging \textit{`relevant'} passages (in line with findings by~\citet{alonso2009can}), followed by `non-relevant' passages. `Somewhat relevant' passages were most difficult to judge as participants in both conditions judged them correctly about half the time. With respect to the time taken to judge (\textbf{III}, Table~\ref{tab:RQ1}), judgements in \textans{} were made significantly faster ($t(47) = -4.93, p < 0.001$) than in \voiceans{}. 

\begin{table*}[!t]
     \centering
    \footnotesize
     \caption{\textbf{RQ2}: Effects of passage length and presentation modality on accuracy of relevance judgements (with  Krippendorff's $\alpha$, Cohen's $\kappa$) and time taken. A \underline{bold} number indicates that the metric for the corresponding presentation modality is significantly more than that for the other modality for the particular passage length. $^{xs, s, m, l, xl}$ indicates significant difference (within the same experimental condition) compared to \XSh{}, \Sh{}, \M{}, \Lo{}, \XLo{} passage lengths. $\star$ indicates equivalence between the two conditions. }
     
     \label{tab:RQ2}
    
     \renewcommand{\arraystretch}{1.3}

     \begin{tabularx}{\linewidth}{>{\hsize=0.1\hsize}Y>{\hsize=1.2\hsize}X||>{\hsize=0.3\hsize}Y|>{\hsize=1.28\hsize}Y|>{\hsize=1.28\hsize}Y|>{\hsize=1.28\hsize}Y|>{\hsize=1.28\hsize}Y|>{\hsize=1.28\hsize}Y}
        & \textbf{Metrics} & \textbf{Mode} & \multicolumn{5}{c}{\textbf{Passage Length}} \\
        & & & \XSh{} & \Sh{} & \M{} & \Lo{} & \XLo{} \\ \hline\hline

        \multirow{4}{*}{\textbf{I}} & \multirow{4}{*}{\textbf{Accuracy (\%)}} & \textans{} & $66.7 ( \pm 19.0)\star$ & $74.5 ( \pm 14.7)$ & $66.5 (\pm 17.5)$ & $61.0 (\pm 18.0)\star $ & $74.0 (\pm 19.0) $ \\
        & & $\alpha$, $\kappa$ & 0.37, 0.57 & 0.51, 0.67 & 0.43, 0.55 & 0.29, 0.50 & 0.44, 0.68 \\
        \cline{3-8}
        & & \voiceans{} & $67.7 (\pm 21.9)\star$ & $64.06 (\pm 15.0)$ & $72.4 (\pm 19.4)$ & $61.5 (\pm 13.9)\star$ & $64.0 (\pm 16.7)$ \\
        & & $\alpha$, $\kappa$ & 0.39, 0.56 & 0.44, 0.49 & 0.49, 0.63 & 0.27, 0.51 & 0.35, 0.48 \\
        \hline

        \multirow{4}{*}{\textbf{II}} & \multirow{4}{*}{\textbf{Maj. Voting Acc. (\%)}} & \textans{} & $75$ & $83$ & $79 $ & $75 $ & $92$ \\
        & & $\kappa$ & 0.73 & 0.81 & 0.74 & 0.73 & 0.91 \\
        \cline{3-8}
        & & \voiceans{} & $79$ & $79$ & $79$ & $67$ & $79$ \\
        & & $\kappa$ & 0.78 & 0.78 & 0.73 & 0.62 & 0.76 \\
        \hline

        \multirow{2}{*}{\textbf{III}} & \multirow{2}{*}{\textbf{Time Taken (sec.)}} & \textans{} & $14.11 ( \pm 6.3) $ &$15.25 ( \pm 7.7)$ &$15.15 (\pm 6.8)$ &  $21.39 (\pm 12.41) $ & $21.86 (\pm 12.7) $ \\
        \cline{3-8}
        & & \voiceans{} & $17.3 (\pm 5.0)^{m,l,xl}$ & $\mathbf{\underline{25.47} (\pm 11.8)}^{xl}$ & $\mathbf{\underline{28.45} (\pm 14.8)}^{xs,xl}$ & $\mathbf{\underline{31.04} (\pm 6.15)}^{xs,xl}$ & $\mathbf{\underline{45.39} (\pm 9.6)}^{xs,s,m,l}$ \\
        
     \end{tabularx}

\end{table*}

In terms of workload measured using NASA-TLX, there was no significant difference in averages between the two cohorts in terms of perceived mental demand, effort, and temporal demand (\textbf{IV}-\textbf{VI}, Table~\ref{tab:RQ1}). The TOST procedure revealed equivalent scores ($p<0.05$) provided by participants for these three items of the NASA-TLX scale. For the other dimensions of NASA-TLX questionnaire, participants in \textans{} reported they felt significantly more frustrated (\textbf{VII},  Table~\ref{tab:RQ1}) while performing the task than those in \voiceans{} ($t(47) = 4.69,p<0.001$). Participants in \voiceans{} also reported significantly higher perceived performance (\textbf{VIII}, Table~\ref{tab:RQ1}) when compared to the former ($t(47) = -3.60, p<0.001$). 

Overall, we found that participants listening to voice passages were equally accurate to their text counterparts.~\citet{vtyurina2020mixed} also observed that the probability of participants to identify the most relevant document was the same for both text and voice conditions. However, the authors implemented a different task design to ours. Their participants were presented with a list of results, and were significantly better at identifying the correct order of relevance when the summaries were presented in text modality. Insofar as to acknowledging the difference in task design, our observations with regards to the accuracy of participants with respect to relevance judgements across modalities are found to be partially in line with those of~\citet{vtyurina2020mixed}. We also observed that \voiceans{} participants perceived a lower or equal workload when compared to those of \textans{}, in contrast to the other study's findings~\cite{vtyurina2020mixed}. This can be attributed to their study setup. Contrary to ours, their presentation modality was a \textit{within-subjects} variable. Our results indicate the proficiency of participants with both modalities for the given design of the task. 
 



\subsection{RQ2: Passage Length}\label{sec:RQ2}

Table~\ref{tab:RQ2} presents results related to \textbf{RQ2}. 
Like modality of presentation, passage length or its interaction with presentation modality did not have a significant effect on the relevance judgement accuracy (comparing rows \textbf{Ia} and \textbf{Ib}, Table~\ref{tab:RQ2}). The TOST procedure revealed that for \XSh{} ($p=0.01$) and \Lo{} ($p=0.001$) passages, judgement accuracy was \textit{equivalent} across both conditions. Aggregating judgements via majority voting increased relevance judgement accuracy across all passage lengths for both text and voice conditions (comparing rows \textbf{Ia}-\textbf{IIa} and \textbf{Ib}-\textbf{IIb}, Table~\ref{tab:RQ2}). However, for \XLo{} passages (\textbf{IIa}-\textbf{IIb}, Table~\ref{tab:RQ2}), the difference in accuracy after majority voting was more than 10\% (with \textans{} being more accurate). We also observed a higher difference in Cohen's $\kappa$ and Krippendorff's $\alpha$ for \XLo{} passages between the \textans{} and \voiceans{} conditions. These results indicated a higher inter-annotator agreement and reliability of judgements for \textans{} compared to participants in \voiceans{} with regards to \XLo{} passages. 

With respect to the time taken for judging, we have already seen (Section~\ref{sec:RQ1}) that presentation modality significantly affected the time to judge. Mixed factorial ANOVA showed that passage length had a significant main effect (F $=21.6, p = 3.3e^{-15})$ on the time taken to assess. A post-hoc test revealed a significant difference in the time taken to judge of the following pairs of passage lengths (with the latter passage length category taking more time): \XSh{}--\M{} ($p=0.02$), \XSh{}--\Lo{} $(p < 0.001)$, \XSh{}--\XLo{}  $(p < 0.001)$, \Sh{}--\XLo $(p < 0.001)$ and \M{}--\XLo $(p = 0.001)$. There was also a significant interaction effect between passage length and presentation modality on the amount of time taken. Pairwise Tukey test revealed that except for \XSh{} passages, judging relevance in \voiceans{} took significantly longer for participants as compared to doing the same in \textans{} (
\textbf{bold numbers}, row \textbf{III}, Table~\ref{tab:RQ3}). In \voiceans{} (\textbf{IIIb}, Table~\ref{tab:RQ3}), it took participants significantly longer to judge relevance, as passages (
audio clips) increased in length.  
Superscripts (in Table~\ref{tab:RQ2}) indicate which pairs of passage length were significantly different in \voiceans{} in terms of time taken per judgement.

In summary, we did not observe a significant difference in relevance judgement accuracy across different passage lengths in both conditions. We observed judging relevance of \XSh{} passages was \textit{equivalent} in terms of accuracy and time taken across both \textans{} and \voiceans{}. However, for \XLo{} passages, relevance judgements in \textans{} were more reliable (indicated by majority voting accuracy, $\alpha$ and $\kappa$ when compared to that in \voiceans{}). There was no clear trend between passage length and assessor agreement observed in contrast to findings from~\cite{chandar2013document}, possibly due to differences in the type of documents assessed. Although it took longer on average to judge a lengthier passage in \textans{}, there was no significant difference in terms of the time taken to judge relevance of different passage lengths (a similar trend as observed in~\cite{konstan1997grouplens,shinoda2012information}). For longer passages, participants in \voiceans{} took significantly longer to judge relevance than in \textans{}. For \XLo{} passages, we found that participants were taking twice as long in \voiceans{} when compared to \textans{}. 
\begin{figure}
\centering

\includegraphics[width=\linewidth]{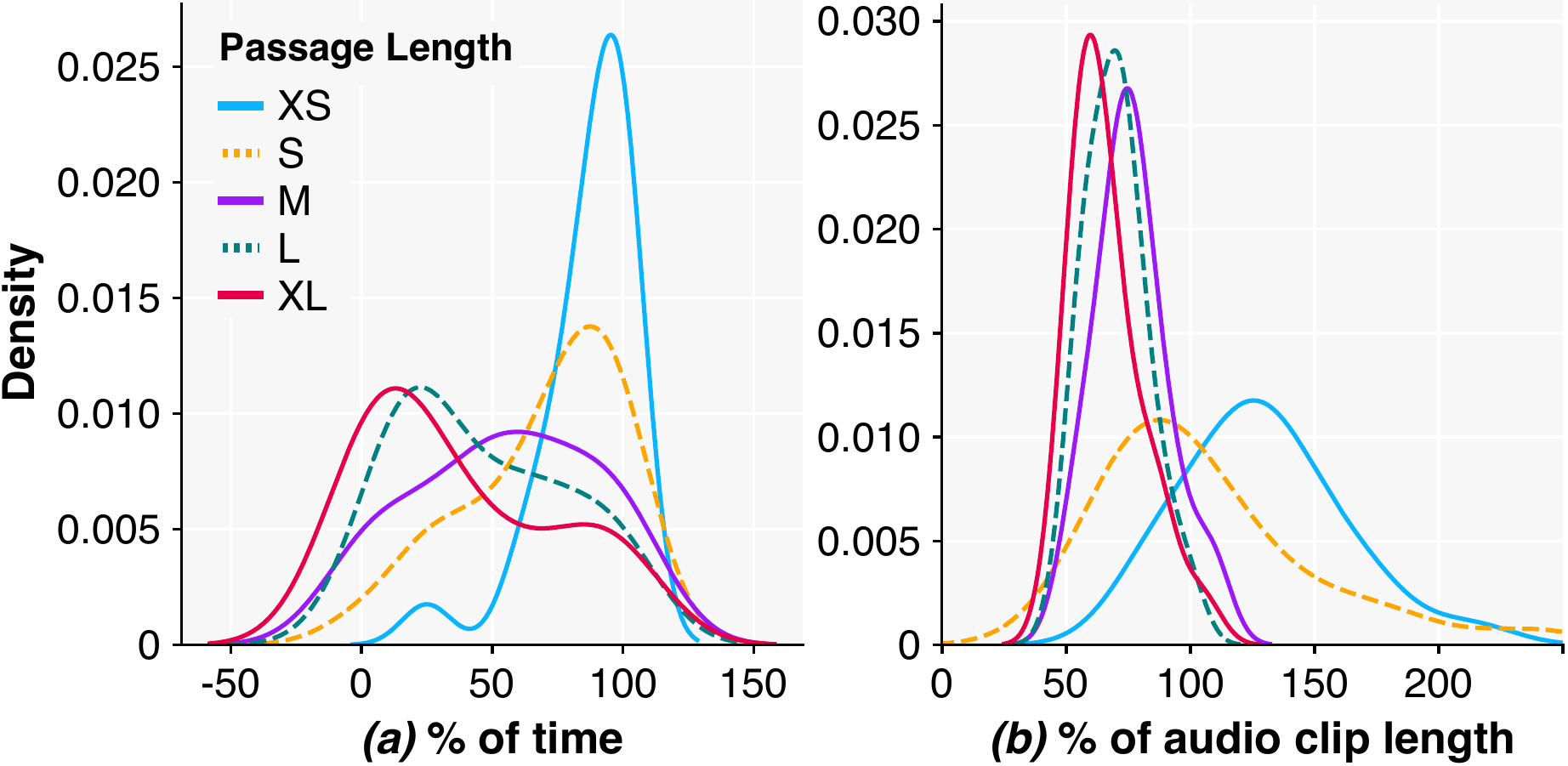}

%
%
\caption{The trend of voice participants judging relevance w.r.t. time taken for passages of various length: \textit{(a)} \% of time participants listened to the entire audio clip; and \textit{(b)} at what point was relevance judged (as a \% of audio clip length).}
\label{fig:voice_long}
\end{figure}
\begin{table*}[!t]
     \centering
    
     \caption{RQ3: Summary of main effects of \textit{Presentation Modality (PM)}, \textit{Working Memory (\WM{})}, \textit{Inhibition (\In{})}, and effects of the interaction of \WM{} and \In{} with PM on accuracy of relevance judgement, time taken, and perceived workload. A \cmark{} indicates significant effect of a 3-way ANOVA test (\boldmath{$p<0.05)$} on the particular dependent variables and \xmark{} indicates no significant effect. }
     \vspace*{-2mm}
     \label{tab:RQ3}
    \footnotesize
     \renewcommand{\arraystretch}{1.3}

    \begin{tabularx}{\linewidth}{X||Y|Y|Y||Y|Y}
         \textbf{} & \textbf{PM} \textit{(Presentation)} & \textbf{\WM{}} \textit{(Working Memory)} & \textbf{\In{}} \textit{(Inhibition)} & 
         \textbf{\WM{}}\boldmath{x}\textbf{PM} &
         \textbf{\In{}}\boldmath{x}\textbf{PM}   \\ \hline\hline
           \textbf{I Accuracy} &   \xmark{} &   \xmark{} &   \xmark{} &     \xmark{} &   \cmark{} (F $=4.89, p = 0.03)$ \\ \hline
         \textbf{II Time Taken (sec.)}  &\cmark{} (F $=22.17, p<0.001)$ & \xmark{} &\xmark{} &\xmark{} & \xmark{} \\ \hline
             \textbf{III Mental Demand} &   \xmark{}  &   \xmark{}  &    \xmark{} &   \xmark{}&   \xmark{}  \\ \hline
         \textbf{IV Effort} & \xmark{} & \xmark{}  &  \xmark{} &  \cmark{} (F $=5.1, p = 0.03)$  & \xmark{}
         \\ \hline
           \textbf{V Temporal Demand} &   \xmark{} &   \cmark{} (F $=7.88, p = 0.01)$ &   \cmark{} (F $=7.39, p = 0.01)$ &   \xmark{}  &   \xmark{} \\ \hline 
         \textbf{VI Frustration}  &\cmark{} (F $=8.36, p=0.008)$ & \xmark{} &\xmark{} &\xmark{} & \xmark{} \\  \hline
             \textbf{VII Performance} &  \cmark{} (F $=5.83, p=0.02)$ &   \xmark{} &   \xmark{} &    \xmark{} &   \xmark{} \\ 
    \end{tabularx}
\end{table*}

\paragraph{Why does it take longer for participants to judge longer passages in the \voiceans{} condition?}  
In order to control for confounding variables, we did not let participants speed up the audio clips, nor did we provide them with a seeker bar to skip ahead. We found evidence that participants moved on to the next Q/P pairing as soon as they were satisfied with their assessment. 
Indeed, they did not wait for the audio clip to finish playing before moving on to the next Q/P pair for longer passages (Figure~\ref{fig:voice_long} (a)). We also let participants mark the relevance of a passage in \voiceans{} only after $50\%$ of the audio clip had been played (Section~\ref{sec:exp_proc}). However, as seen from Figure~\ref{fig:voice_long} (b), participants 
took longer to judge relevance (rather than right at the $50\%$ mark). For \XLo{} passages, it was at the $66\%$ of the audio clip on average. 
This suggests that it indeed took more time for participants in \voiceans{} compared to \textans{} to assimilate the information and come to a judgement decision for longer passages. 


\subsection{RQ3: Assessor Cognitive Abilities}\label{sec:RQ3}

Table~\ref{tab:RQ3} contains the results for our third research question. Here, \cmark{} indicates a significant effect ($p<0.05)$ on the particular dependent variable, and \xmark{} indicates no significant effect.

None of the independent variables---modality of passage presentation (\textbf{PM}), working memory (\WM{}), and inhibition (\In{})---had a significant main effect on judgement accuracy. The interaction between the \In{} of participants and presentation modality (\In{} x \textbf{PM}) had a significant effect on the accuracy (F $=4.89, p = 0.03)$.  Pairwise Tukey test revealed that in \voiceans{} participants with higher \In{} performed significantly better than those with lower \In{} ($70.5\pm 7.2$\% vs. $59.5\pm 4.8$ \%).
The post-hoc test $(p=0.01)$ also revealed participants with low \In{} performed significantly better in \textans{} than those in \voiceans{} ($70.0\pm 9.5$ \% vs. $59.5\pm 4.8$ \%). We found significant main effects of \textbf{PM} on the time taken to judge relevance (F $=22.17, p<0.001)$, reaffirming findings from Section~\ref{sec:RQ1} and Section~\ref{sec:RQ2}. 

With respect to the perceived workload, working memory had significant main effects on perceived temporal demand (F $=7.88, p = 0.01)$. A post-hoc test ($p<0.001)$ revealed that participants with high \WM{}  reported significantly less temporal demand as compared to those with low \WM{} ($2.5\pm 1.3$ vs. $4.6\pm 1.7$ respectively). \In{} also had significant main effects on perceived temporal demand (F $=7.4, p = 0.01)$. A post-hoc test ($p<0.001)$ revealed that participants with high \In{} reported significantly less temporal demand as compared to those with low \In{} ($2.74\pm 1.4$ vs. $4.59\pm 1.9$, respectively). Presentation modality had significant main effects on perceived frustration (F $=8.36, p = 0.008)$ and performance (F $=5.83, p = 0.02)$---confirming observations from Section~\ref{sec:RQ1}---with participants in \voiceans{} reporting a lower workload. Lastly, the interaction between \WM{} and presentation modality (\WM{} x \textbf{PM}) had a significant effect on perceived effort for the task (F $=5.1, p = 0.03)$. Post-hoc tests revealed that participants with high \WM{}  felt that judging using \textans{} required significantly more effort when compared to those in \voiceans{} ($p=0.001)$. 

In summary, we found that \In{} is a more important trait than \WM{}, specifically for relevance judgement accuracy in the \voiceans{} modality. Low \In{} participants in the \voiceans{} condition were less accurate---since we \textit{did not control for the audio device of the participants}, and consequently not for the background noise they were subjected to, low \In{} participants in \voiceans{} were less effective in focusing on the passages while judging relevance~\cite{stenback2016speech,stenback2021contribution}. We leave exploring the effect of background noise as future work. In our study, the interplay between cognitive abilities and modality of presentation on perceived workload had different effects. High \In{} and \WM{} participants felt less temporal demand. High \WM{} in \textans{} felt more perceived effort compared to those in \voiceans{}. Our results imply that we should design tasks for collecting relevance assessments to match the preference and abilities of crowdworkers~\cite{organisciak2014crowd,alsayasneh2017personalized}. 



 \section{Conclusions}
\label{sec:conclus}

We explored the feasibility of using \voiceans{} as a modality to collect relevance judgements of query-passage pairs. We investigated the effect of passage length and the cognitive abilities of participants on judgement accuracy, the time taken, and perceived workload. 

\vspace*{1mm}\noindent\textbf{\rqOne{}} On average, the relevance judgement accuracy was equivalent across both \textans{} and \voiceans{}. Participants also perceived equal or less workload in \voiceans{} when compared to \textans{}.

\vspace*{1mm}\noindent\textbf{\rqTwo{}} For \XSh{} passages, the performance and time taken for relevance judgements was \textit{equivalent} between both \voiceans{} and \textans{}. As passages increased in length, it took participants significantly longer to make relevance judgements in the \voiceans{} condition; for \XLo{} passages \voiceans{}, participants took twice as much time and the judgements were less reliable compared to \textans{}.
       
\vspace*{1mm}\noindent\textbf{\rqThree{}} Inhibition impacted the relevance judgement accuracy in the \voiceans{} condition---participants with higher inhibition were significantly more accurate than those with lower inhibition.

\vspace*{1mm}

Our results from \rqOne{} suggest that we can leverage the voice modality for this task. \rqTwo{} points to the possibility of designing hybrid tasks, where we can use the voice modality for judging shorter passages and text for longer passages. The results of \rqThree{} showed that selecting the right participants for the relevance judgement task is important. We should be mindful to personalise the task to match the preference and abilities of crowdworkers~\cite{organisciak2014crowd,alsayasneh2017personalized}. 

There are several open questions for future work. We did not provide participants with the option to speed-up voice passages---\textit{does letting them speed-up or skip passage parts reduce time for longer passages without reducing accuracy?} We also did not test the limit of length---\textit{how long can documents be for equal accuracy in the text and voice modality?} 
Future work should also explore mobile devices for playing voice passages---\textit{can we collect relevance judgements by offering more flexibility to crowdworkers?} 
Lastly, since asking to provide rationales for judgements has been shown to improve relevance judgement accuracy of crowdworkers in the text modality~\cite{kutlu2020annotator}, exploring the effects of rationale in voice-based relevance judgements should be a worthwhile endeavour.

\bibliographystyle{ACM-Reference-Format}
\bibliography{references}
\end{document}